\documentclass[prl,twocolumn,floatfix,showpacs,showkeywords]{revtex4}

\usepackage{amssymb}
\usepackage{amsmath}
\usepackage{amsfonts}
\usepackage{bm}
\usepackage{graphicx}

\newcommand{\br}{{\bf r}}
\newcommand{\bQ}{{\bf Q}}
\newcommand{\bn}{{\bf n}}
\newcommand{\bM}{{\bf M}}

\newcommand{\bR}{{\bf R}}
\newcommand{\bk}{{\bf k}}
\newcommand{\bh}{{\bf h}}
\newcommand{\bbf}{{\bf f}}
\newcommand{\ud}{\textrm{d}}

\newcommand{\nn}{\nonumber}

\begin{document}

\title{Quantum order-by-disorder near criticality and the secret of partial order in MnSi}

\author{Frank Kr\"uger$^1$}
\author{Una Karahasanovic$^1$} 
\author{Andrew G. Green$^2$}
\affiliation{$^1$SUPA, School of Physics and Astronomy, University of St. Andrews, KY16 9SS, United Kingdom\\
$^2$London Centre for Nanotechnology and Department of Physics and Astronomy, University College London, London WC1E 6BT, United Kingdom}

\date{\today}

\begin{abstract}
 The vicinity of quantum phase transitions has proven fertile ground in the search for new quantum phases. We propose a physically 
 motivated and unifying description of phase reconstruction near metallic quantum-critical points using the idea of quantum order-by-disorder. 
 Certain deformations of the Fermi surface associated with the onset of competing order enhance the phase space available for low-energy, particle-hole
 fluctuations and self-consistently lower the free energy. Applying the notion of quantum order-by-disorder to the itinerant helimagnet MnSi, we show that 
 near to the quantum critical point, fluctuations lead to an increase of the spiral ordering wave vector and a reorientation away from 
 the lattice favored directions. The magnetic ordering pattern in this fluctuation-driven phase is found to be in excellent agreement with the neutron 
 scattering data in the partially ordered phase of MnSi.
\end{abstract}


\maketitle

For a long time MnSi was considered a textbook example of a Fermi liquid developing ferromagnetic order at low temperatures where small spin-orbit (SO) 
coupling leads to a long-wavelength helical modulation of the magnetic order \cite{Bak+80,Ishikawa+76}. This view changed when experiments found 
a radical departure from conventional metallic physics upon suppression of  magnetism under hydrostatic pressure. The occurrence of this 
behavior over a wide pressure range, not just close to the transition \cite{Pfleiderer+01}, suggests a new phase of matter. A clue to its nature 
was provided by neutron scattering \cite{Pfleiderer+04}, which revealed an unusual partially ordered phase, not apparent in susceptibility or resistivity. 
While the helimagnetic order is characterized by resolution-limited peaks corresponding to spiral wavevectors $Q \sim 0.037 \AA^{-1}$ parallel to the $[111]$ 
directions, in the partially ordered phase the scattering signal spreads diffusely over a sphere of radius $Q\sim 0.043 \AA^{-1}$, weakly favoring the $[110]$ 
directions and anti-favoring the $[111]$ and $[100]$ directions. This is suggestive of spirals that fluctuate in direction, avoiding  pinning by the 
lattice. 

In itinerant ferromagnets without SO coupling, it has been understood for some time that quantum fluctuations can drive the magnetic transition
 first order at low temperatures \cite{Abrikosov+58,Conduit+09b}. The same physics has been revealed more recently in non-analytic corrections 
 \cite{Belitz+97} to the Hertz-Millis theory \cite{Hertz76} of ferromagnetic quantum criticality. This latter
approach has been extended to systems with small SO coupling \cite{Vojta+01+Schmalian+04}, explaining the weak first order behavior seen in early experiments
 on MnSi \cite{Pfleiderer+97}. 
Various itinerant ferromagnets display not only weak first-order behavior, but also the emergence of unusual phases in the vicinity of putative 
quantum critical points \cite{Uhlarz+04,Borzi+07}, suggesting that the avoidance of naked quantum critical points represents 
a generic principle \cite{Laughlin+01}. In homogeneous ferromagnets the quantum fluctuations that drive the transition first order also
stabilize an inhomogeneous spiral phase, preempting the first-order transition \cite{Belitz+97,Conduit+09}. 

Formulation of these ideas in the form of quantum order-by-disorder \cite{Conduit+09} (or Coleman-Weinberg) provides an attractive physical interpretation. 
Moreover, it guides one to a simple route through the calculations. The central idea is that certain deformations of the Fermi surface enhance the phase space 
available for low-energy quantum fluctuations and so self-consistently lower the free energy. This mechanism is familiar in condensed matter \cite{Mila+91}. Unusually, 
here it is driven by fermionic rather than bosonic fluctuations.  The framework suggests direct connections to various experimental probes.  
Fermi-surface deformations and reconstructions associated with the onset of the competing order can be observed in photoemission and recently,
the measurement of entropic landscapes has  proven a revealing probe of phase reconstruction near quantum critical points \cite{Borzi+07}.

The utility of the quantum order-by-disorder approach is exemplified in its application to MnSi. We argue that at low temperatures near to the putative quantum 
critical point, the spiral wavevector shifts away from its amplitude and direction in the ordered phase in order to open up and benefit energetically from extra phase 
space for low-energy particle-hole excitations. This drives an instability towards a phase characterized by spirals that fluctuate about 
new directions, with an angular dependence as measured by neutron scattering \cite{Pfleiderer+04}. 

The helimagnetic phase of MnSi is well described by a mean-field theory combining Stoner ferromagnetism with SO coupling resulting from the lack of 
inversion symmetry of the cubic B20 crystal structure. This can be captured phenomenologically by a Ginzburg-Landau free-energy density 
${\cal F}= (\nabla\bM)^2+r_0\bM^2+2D\bM\cdot(\nabla\times\bM)+\ldots$, with $D$ the Dzyaloshinskii-Moriya interaction. ${\cal F}$ is minimized 
by planar spirals $\bM(\br)=M\left[\bn_x \cos(\bQ\br)+\bn_y \sin(\bQ\br)\right]$ with propagation vector $\bQ=D\bn_z$ (Fig. 1a).  
While ${\cal F}$ is invariant under continuous rotations of the spiral, higher order spin-orbit  terms lead to small corrections that reduce the symmetry 
to the point-group of the crystal. For MnSi, the leading anisotropy is given by \cite{Bak+80}  
$\delta {\cal F}=\lambda \sum_i (\partial_i M_i)^2$ ($i=a,b,c$ labels crystal axes) which, for $\lambda<0$, 
leads to the experimentally observed pinning in the [111] direction \cite{Ishikawa+76}. 

The coefficients of the Landau function may be determined microscopically by an expansion of the free energy
\begin{equation}
F_\textrm{mf}  = UM^2-T\sum_{\bk,\alpha=\pm}\ln(1+e^{-(\epsilon^{\alpha}_\bk-\mu)/T}),
\label{free_energy_mf}
\end{equation}
where $U$ is the contact interaction between electrons, $\mu$ the chemical potential. $\epsilon^{\pm}_\bk$ are the mean-field electron dispersions in the presence of a planar spiral magnetization
obtained from the microscopic Hamiltonian 
\begin{eqnarray}
H  & = &   \sum_{\bk,\nu=\uparrow,\downarrow} k^2 c^\dagger_{\bk\nu}c_{\bk\nu}+U\sum_\br \hat{n}_{\br\uparrow}\hat{n}_{\br\downarrow}\nn\\
& & -\frac 12 \sum_{\bk,\nu,\nu'}\bh(\bk)\cdot \bm{\sigma}_{\nu,\nu'}c^\dagger_{\bk\nu}c_{\bk\nu'},
\label{hamiltonian}
\end{eqnarray}
with quadratic electron dispersion, contact repulsion and SO interaction $\bh(\bk)$. $c^\dagger_{\bk\nu}$ and $c_{\bk\nu}$ denote fermion creation and annihilation operators, $\hat{n}_{\br\nu}$ the occupation operator in real space, 
and $\bm{\sigma}_{\nu,\nu'}$ the vector of Pauli matrices. The SO coupling is a consequence of the lack of inversion symmetry of the crystal structure
of MnSi and, in the crystal axes basis, of the form $\bh_c = \alpha [k_a,k_b,k_c]+\tilde \alpha [k_a(k_b^2-k_c^2),k_b(k_c^2-k_a^2), k_c(k_a^2-k_b^2)]$ \cite{Frigeri+04}.
Note that in the Hamiltonian (\ref{hamiltonian}) only the higher order SO coupling, $\tilde\alpha$, breaks the rotational symmetry and 
is therefore responsible for the directional dependencies of both the helimagnet and the partially ordered phase.

The mean-field decoupling of the interaction takes the form 
$H_U\approx U\sum_{\br,\nu,\nu'} \bM(\br)\cdot  \bm{\sigma}_{\nu,\nu'}c^\dagger_{\br\nu}c_{\br\nu'}=UM\int_\bk(c^\dagger_{\bk+\bQ/2\uparrow}c_{\bk-\bQ/2\downarrow}
+\textrm{h.c.})$. We have chosen the propagation vector $\bQ=Q\bn_z=Q(\sin\theta\cos\phi\bn_a+\sin\theta\sin\phi\bn_b+\cos\theta\bn_c)$ as quantization axis and so rotate the spin-orbit interaction to the spiral basis; $\bh(\bk)=\bR_{\phi,\theta}\bh_c(\bR_{\phi,\theta}^T\bk)=\alpha\bk+\tilde \alpha\bbf(\bk,\phi,\theta)$ where 
$\bR_{\phi,\theta}\bn_c=\bn_z$ (see Fig.~1b). The components of $\bbf$ are lengthy third-order polynomials in $k_x$, $k_y$, $k_z$ with coefficients that depend on the 
direction of $\bQ$. To leading order (linear coupling to $Q$, corresponding to the Dzyaloshinskii-Moriya interaction) only $h_z$ contributes, 
yielding dispersions
\begin{eqnarray}
\epsilon^\pm(\bk) & = & k^2-\tilde \alpha Q\Gamma_1(\bk,\phi,\theta)\\
 & & \pm \sqrt{\left[k_z(Q-\alpha)-\tilde \alpha Q^2\Gamma_2(\bk,\phi,\theta)   \right]^2+(UM)^2 },\nonumber
\label{dispersion}
\end{eqnarray}
where $Q\Gamma_1=[f_z(\bk+\frac{\bQ}{2})- f_z(\bk-\frac{\bQ}{2})]/2$ and $Q^2\Gamma_2=[f_z(\bk+\frac{\bQ}{2})+f_z(\bk-\frac{\bQ}{2})]/2$.
The isotropic SO coupling leads to the Dzyaloshinskii-Moriya interaction with $D=\alpha$. 
As  shown later, the leading, mean-field anisotropy  is indeed of the  form $\delta {\cal F}=\lambda \sum_i (\partial_iM_i)^2$ 
with $\lambda\sim \tilde{\alpha}^2$.  

In order to analyze the stabilization of competing ground states by quantum fluctuations, we include the leading fluctuation corrections which in self-consistent second 
order perturbation theory are given by \cite{Conduit+09}
\begin{equation}
F_\textrm{fl}=-2U^2\sum_{\bk_1+\bk_2=\bk_3+\bk_4}\frac{n(\epsilon^+_{\bk_1})n(\epsilon^-_{\bk_2})[n(\epsilon^+_{\bk_3})+n(\epsilon^-_{\bk_4})]}{\epsilon^+_{\bk_1}+
\epsilon^-_{\bk_2}-\epsilon^+_{\bk_3}-\epsilon^-_{\bk_4}},
\label{free_energy_fl}
\end{equation}
with $n(\epsilon)=(e^{(\epsilon-\mu)/T}+1)^{-1}$ the Fermi function. 

\begin{figure}[t]
\includegraphics[width=0.75\linewidth]{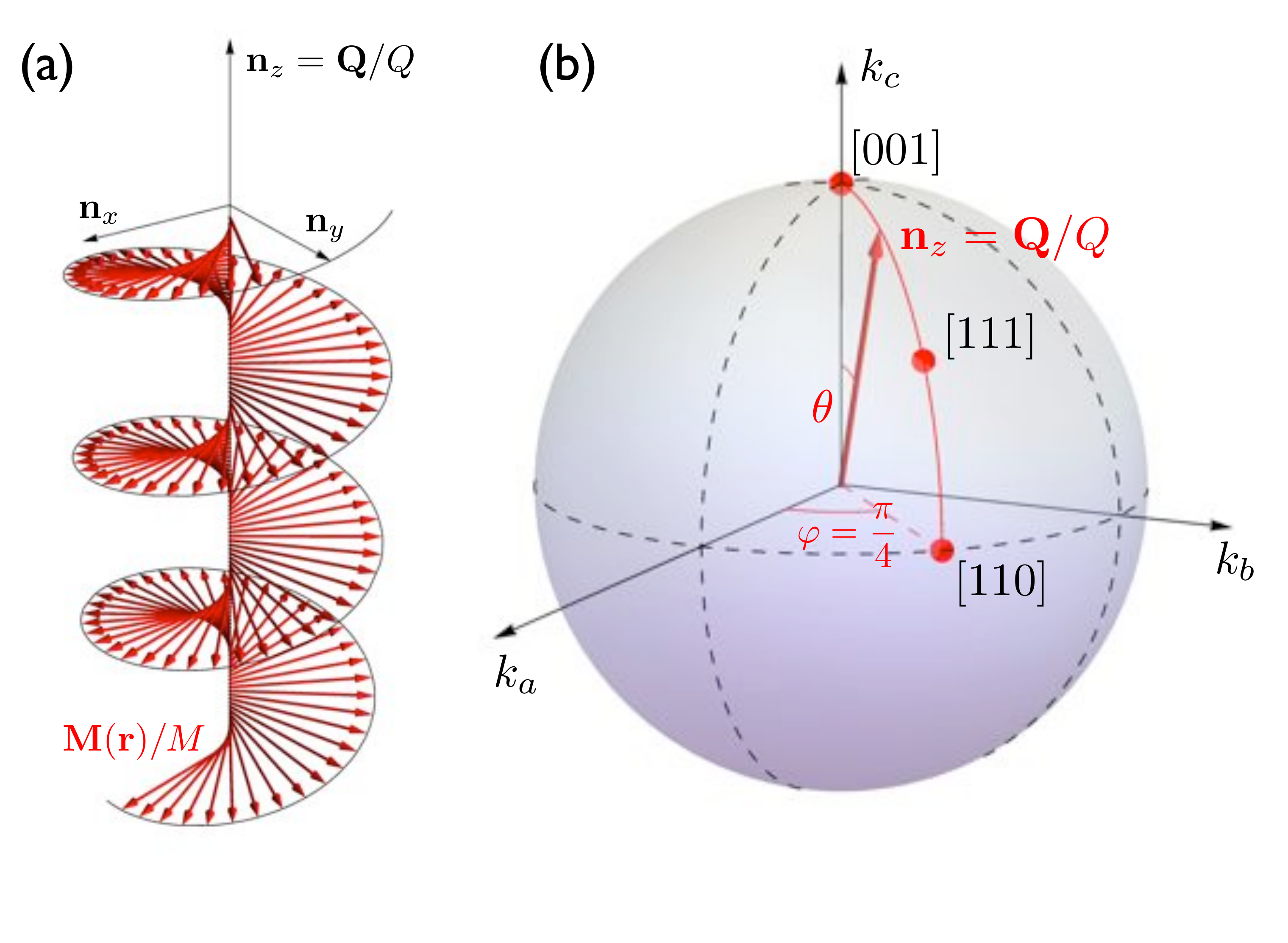}
\caption{(color online) (a) Planar spiral $\bM(\br)=M\left[\bn_x \cos(\bQ\br)+\bn_y \sin(\bQ\br)\right]$ 
with propagation vector $\bQ=Q\bn_z$. The magnetic moments of size $M$ are confined to a plane spanned by $\bn_x, \bn_x\perp \bQ$. (b) Due 
to higher order SO terms in the cubic B20 crystal structure of MnSi, the free energy depends on the direction of 
$\bQ=Q(\sin\theta\cos\phi\bn_a+\sin\theta\sin\phi\bn_b+\cos\theta\bn_c)$ with respect to the crystal axes $\bn_a$, $\bn_b$ and $\bn_c$.}
\label{fig.1}
\end{figure}

Since the anisotropic SO coupling, $\tilde \alpha$, is an order of magnitude smaller than $\alpha$, it has a negligible effect upon the location of the 
phase boundaries and the pitch $Q$ \cite{Belitz+06}. Therefore, we calculate the magnetic phase diagram for $\tilde \alpha=0$ and then analyze 
the directional dependencies.  For $\tilde \alpha=0$, the mean-field dispersion is identical 
to that without SO coupling if we replace $Q$ by $\tilde{Q}=Q-\alpha$. Therefore the phase diagram is the same as that of the homogeneous systems with 
the appropriate shift of $Q$. The homogeneous system shows a ferromagnet to paramagnetic transition with a fluctuation induced spiral phase forming around the putative quantum critical point. In the presence of SO coupling, a helimagnet to paramagnetic transition is obtained with a small region of fluctuation induced partial order 
around the critical point. 

Restricting our consideration to planar spiral configurations, we seek an expansion of the fluctuation-corrected free energy of the form
$F=(a_0 +a_2 \tilde{Q}^2+a_4\tilde{Q}^4)M^2+(b_0 +b_2\tilde{Q}^2)M^4+c_0 M^6$, where the coefficients $a_0$, $a_2$, {\it etc.} are functions of $U$, $T$ and 
$\mu$. This expansion is controlled around a finite temperature tricritical point (that will be revealed below) at which $M=0$. 

Determination of the phase diagram is simplified by relationships between the expansion coefficients: At low temperatures, derivatives of the Fermi functions are 
sharply peaked at the Fermi surface leading to the proportionalities $a_2\approx 2\langle \hat{k}_z^2\rangle b_0$, $a_4\approx 3\langle \hat{k}_z^4\rangle c_0$ and  $b_2\approx 3\langle \hat{k}_z^2\rangle c_0$ ($\hat{k}_z=k_z/k$) with angular averages $\langle   \hat{k}_z^{2n}\rangle=1/(2n+1)$. The remaining coefficients $a_0$, $b_0$ and $c_0$ 
for $\tilde Q=0$ can be computed numerically over the whole temperature range and analytically at low temperatures  \cite{Una_unpublished}. 
In the limit $T\to 0$, we obtain $a_0^\textrm{mf}=u-\frac{2\sqrt{2}}{(2\pi)^2} u^2$, 
$b_0^\textrm{mf}=\frac{\sqrt{2}}{4!(2\pi)^2}u^4$ and $c_0^\textrm{mf}=\frac{15\sqrt{2}}{4\cdot6!(2\pi)^2}  u^6$  for the mean-field coefficients and 
$a_0^\textrm{fl}\simeq -\lambda (1+2\ln2)u^4$ and $b_0^\textrm{fl}\simeq \lambda (1+\ln t)u^6$ for the fluctuation corrections, with dimensionless units $u=U/\mu$ and $t=T/\mu$ and where $\lambda=\frac{16\sqrt{2}}{3 (2\pi)^6}$.

\begin{figure}[t]
\includegraphics[width=0.95\linewidth]{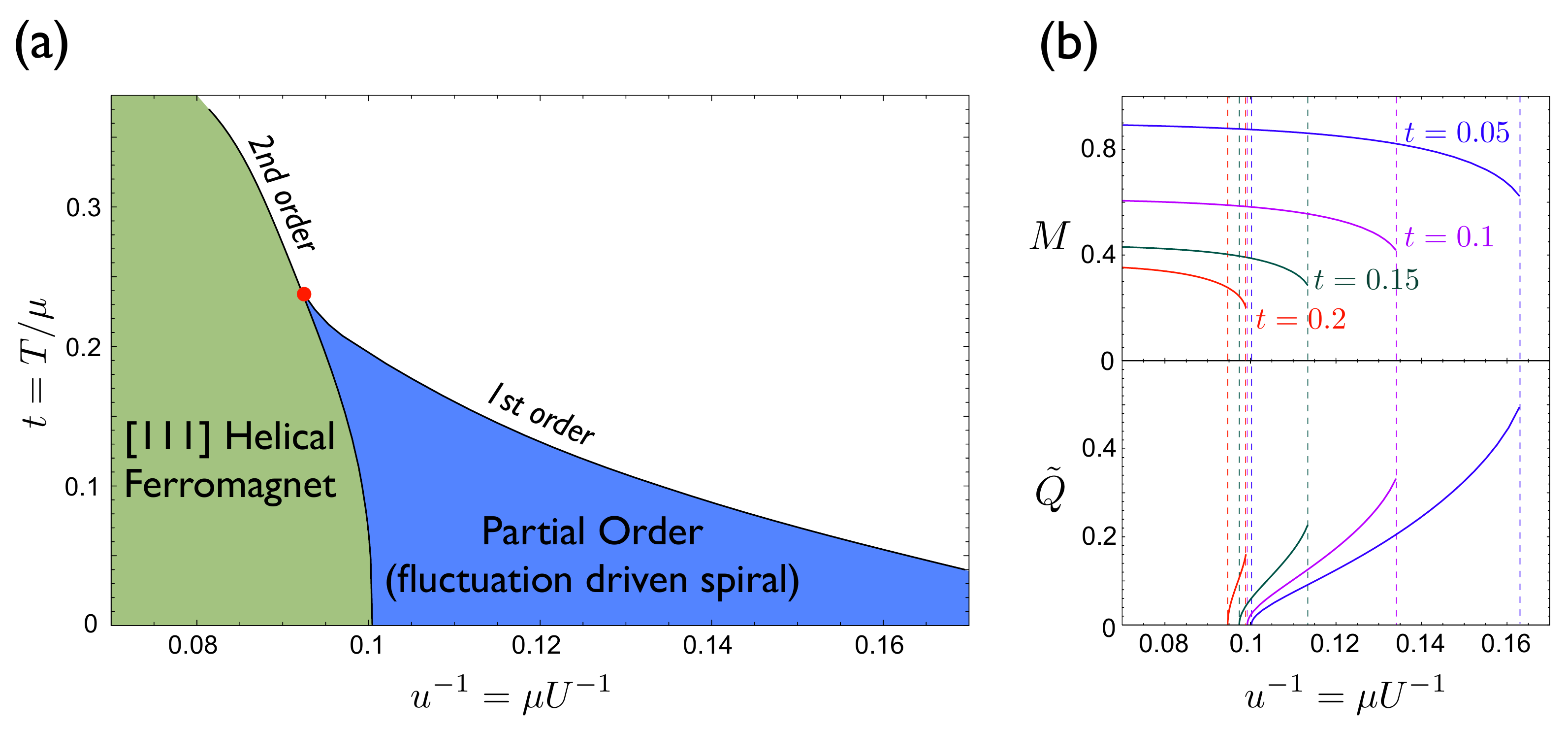}
\caption{(color online) (a) Magnetic phase diagram for weak SO interaction $\alpha$ as a function of inverse interaction $u^{-1}=\mu/U$ and temperature $t=T/\mu$.  
For strong interactions, the system develops helimagnetic order of pitch $Q_0=\alpha$. The transition to the 
paramagnet is continuous. For smaller $u$, quantum fluctuations stabilize a different spiral ground state which emerges from the tricritical point shown in 
red. The higher-order SO term $\tilde \alpha$, which is responsible for the directional dependencies, has a negligible effect upon the location of the phase boundaries 
and the pitch $Q$ \cite{Belitz+06}. (b) Magnetization $M$ and relative pitch $\tilde{Q}=Q-Q_0$ as a function of $u^{-1}$ for $t=0.05, 0.1, 0.15, 0.2$. 
At the transition between the helimagnet and the partially ordered phase, $\tilde{Q}\sim (u^{-1}-u^{-1}_c)^{1/2}$ while $M$ is continuous. The transition to the
paramagnet  is characterized by a first-order jump in $M$.}
\label{fig.2}
\end{figure}

The phase diagram  is shown in Fig. 2 as a function of $t$ and $u$ --- we expect this to reproduce 
the topology of the temperature-pressure phase diagram. A second order phase transition between the helimagnet  and the paramagnet occurs along 
the curve $a_0=0$.  A crucial factor determining  the phase diagram is that quantum fluctuations make a $\ln t$ contribution  to $b_0$ \cite{note}, which becomes 
negative below  $t^* = 0.24$ as a result. On its own, this would imply that the  transition turns first order. In combination 
with the concurrent change in sign of $a_2\approx 2 b_0/3$, this signifies the onset of fluctuation-driven re-orientation of the spiral order and an increase 
in the spiral pitch. The transition between the fluctuation-driven spiral and the paramagnet along the curve $a_0c_0=\frac{17}{63}b_0^2$ is first order
and characterized by a jump in magnetization (Fig.~2b). 
The transition between the two spiral phases ($a_0=0$ line at $t<t_c$) is a Lifshitz line where $Q$ increases continuously. Deviations from a $k^2$ dispersion render the transition weakly first order \cite{Una_unpublished}.
Experimentally, a small increase in $Q$ is indeed observed \cite{Pfleiderer+04}.

The dependence of the free energy upon the direction of ${\bf Q}$ results from the anisotropic higher order SO coupling $\tilde \alpha$ and the corresponding 
deformations $\Gamma_{1/2}$ of the electron dispersion. At low $T$, the resulting free-energy contributions are given by $\delta F_1  =  \frac{15}{8}\tilde \alpha^2 a_0 \langle \Gamma_1^2\rangle Q^2M^2$ and $\delta F_2  =  2\tilde \alpha^2 b_0 \langle \Gamma_2^2 \rangle Q^4M^2$, where $\langle\ldots\rangle$ denotes an angular average over the direction of $\bk$. $\delta F_1$ is dominant in the ordered helimagnet and precisely of the form $\lambda\int\ud^3\br \sum_\alpha (\partial_\alpha M_\alpha)^2$ \cite{Bak+80} 
when evaluated for planar spirals, where $\lambda = -\frac{3}{20}a_0\tilde\alpha^2$. It is minimized for spirals pointing along the $[111]$ directions. Although  
$\delta F_2$  is of higher order in the SO coupling $\alpha\simeq Q$ it dominates the directional dependence in the partially ordered phase since it is proportional to $b_0$ which diverges 
logarithmically at low $T$. The angular average yields $\langle\Gamma_2^2\rangle=g(\phi,\theta)$ with 
\begin{eqnarray}
g(\phi,\theta) & = &  \frac{1}{16}\sin(2\phi)^2\left[\sin(\theta)-3\sin(3\theta)  \right]^2\nn\\
& & +\cos(2\phi)^2\sin(2\theta)^2
\end{eqnarray}
which has maxima in the $[110]$ directions and vanishes along $[111]$ and $[100]$. Evaluating $\delta F= \delta F_1+\delta F_2$ 
for different spiral orientations and setting
$\delta F_{[111]}=\delta F_{[110]}$ we obtain the condition $a_0=\frac 43 b_0 \alpha^2$ for the directional change. 
Up to order $\alpha$ to which we obtained the phase diagram, this coincides with the phase boundary between the spiral phases given by $a_0=0$.

Going beyond  the Ginzburg-Landau analysis and allowing for thermal fluctuations, the system forms domains in which the spirals are oriented around 
one of the principle directions. Within these domains, the system forms a statistical mechanical ensemble of spirals with fluctuating and spatially modulated 
orientation about this principle direction. Since the partially ordered phase is driven by fluctuations, the angular distribution is expected to be larger and the timescale of
fluctuations to be shorter than in the helimagnetic phase.

Allowing for slow spatial modulation of the spiral orientation, we find a contribution to the free energy density given by
$F= \frac{2}{3} M^2 |b_0| (\nabla \hat {\bf Q})^2$. Evaluating the fluctuations in orientation using this and the previously calculated anisotropy, we obtain the same 
form as using the Boltzmann weight  $\exp [- V_\textrm{eff} \delta F_2/T]$ with $V_\textrm{eff}(T) \simeq (M c_0^2 \tilde \alpha^{-2} |b_0|^{-3/2}T^{-1/2}) 
 \exp[\tilde \alpha^2 M^2 |b_0|^3/(c_0^2T)]$. 
 At very low temperatures,  $V_\textrm{eff}\rightarrow \infty $ 
 and fluctuations in orientation become small. 
 In Fig.~3,  this weight is shown along a great circle through the high-symmetry points and as intensity plot over the momentum sphere documenting the excellent 
 agreement with the neutron-scattering data \cite{Pfleiderer+04}. Since $\delta F_1$ and $\delta F_2$ are proportional to $M^2$, and $V_\textrm{eff}$ is proportional to 
 $M$, all fall towards zero along the helimagnet to paramagnet transition and are significantly reduced in the partially ordered phase (Fig.~3b). We anticipate neutron 
 scattering peaks that are much more diffuse in angle in both these regions of the phase diagram.

The timescales of fluctuations in the direction of $\bQ$ are harder to calculate. Presumably they can be approximated by a Fermi Golden Rule calculation of the rate for a two-step 
process where the electron fluid absorbs a low-energy phonon whilst producing a particle-hole pair, followed by recombination of the particle-hole pair in a state with rotated $\bQ$ and 
emission of a phonon with shifted energy. The enhanced density of low-energy particle-hole states that drive the formation of the partially ordered phase  would then 
naturally lead to faster
angular fluctuations. This expectation is consistent with the fact that the partially ordered phase is not visible in NMR - a probe that averages the spin 
configuration over a longer timescale than neutron scattering.

\begin{figure}[t]
\includegraphics[width=0.9\linewidth]{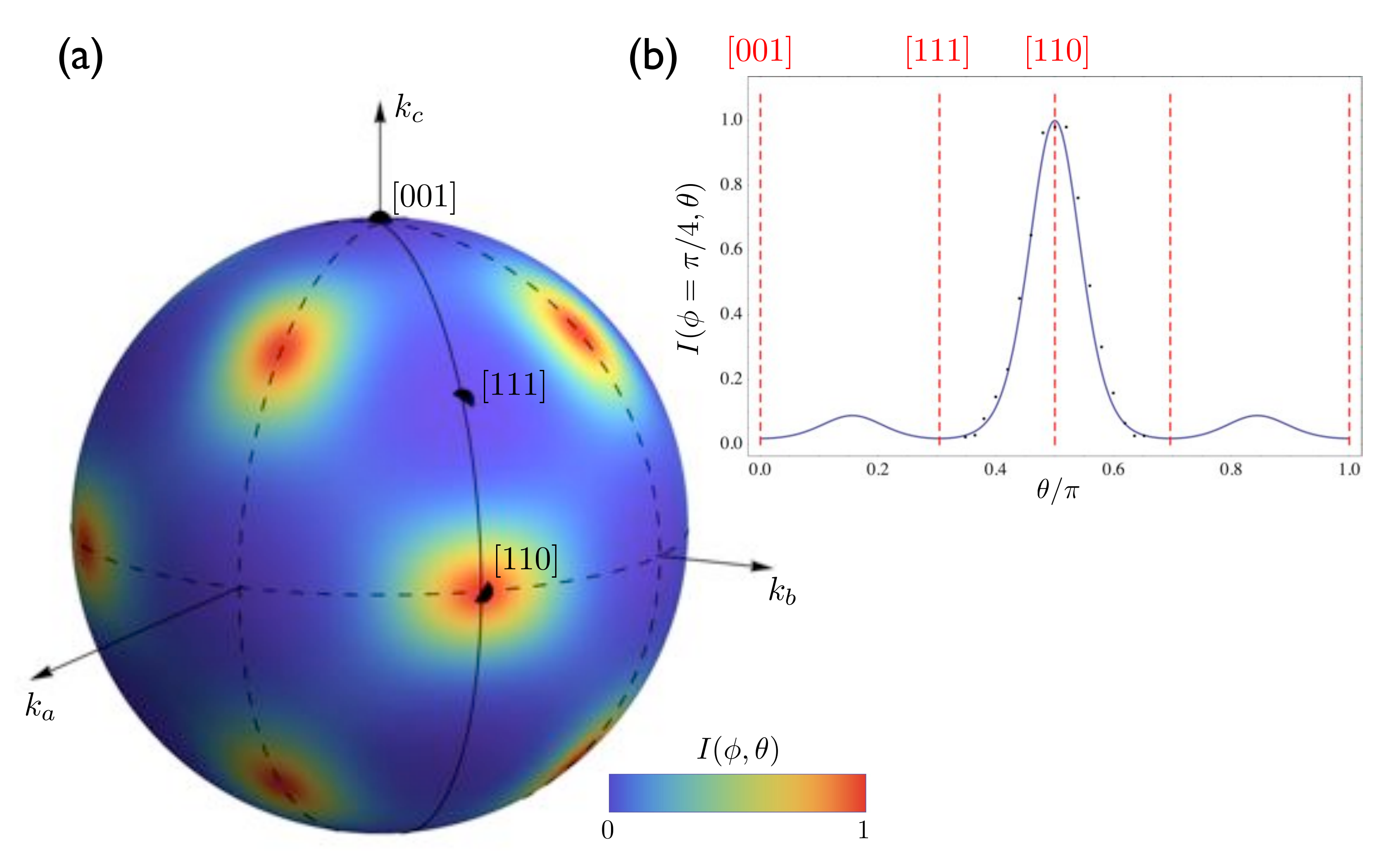}
\caption{(color online)  (a) Boltzmann weight 
$I(\phi,\theta)=I_0\exp [- V_\textrm{eff} \delta F_2 /T]=I_0\exp [c g(\phi,\theta)]$ of the term $\delta F_2$ which dominates the directional dependenve in the partially 
ordered phase. The overall prefactor $c$ enters as the only free fitting parameter. Fluctuations favor spirals along [110] and equivalent directions while 
almost no intensity is found along the [111] directions of the helimagnet.  (b) $I(\phi,\theta)$ with $c=4$ along a great circle connecting the high-symmetry directions in quantitative 
comparison with the experimental data \cite{Pfleiderer+04}.}
\end{figure}

{\it Discussion and Conclusions}:
Quantum order-by-disorder provides a natural explanation of the partially ordered phase of MnSi. In particular, the angular distribution identified from 
neutron scattering is reproduced;  the leading anisotropy of the SO coupling responsible for the [111] ordering in the helimagnetic phase automatically 
leads to [111] (and [100]) directions being anti-favored by fluctuations. 
This had proven problematic in previous analyses --- 
it was the motivation for an alternative suggestion of skyrmion-like ordering \cite{Binz+06}. Skyrmion-like spin crystals (formed by the 
superposition of spirals with axes pointing in different directions) seem necessary to understand features in Hall effect seen at finite magnetic field \cite{Neubauer+98} 
and are compatible with our analysis. However, they are unnecessary to describe the phase boundaries and anisotropy of the partially ordered phase. 

Several measurements would  distinguish between spirals and skyrmions:
i. Annealing in a magnetic field applied along one of the principle directions would align the spiral domains. 
Neutron scattering would show an 
intensity imbalance of the scattering peaks  - an effect not expected for skyrmions;
ii. Annealed samples should also show the anisotropic transport found upon phase reconstruction near to other quantum critical points \cite{Borzi+07}; 
iii. The Lifshitz-like transition between helimagnet and partial order 
is expected to show a diffuse spreading 
of scattering over the whole sphere along the transition line.

The 
unusual, non-Fermi liquid $T^{3/2}$ resistivity \cite{Pfleiderer+01} remains puzzling. It is suspected that 
the effects of disorder upon the partially ordered 
phase provide the key to understanding it \cite{Kirkpatrick+10}.  The consequences of this within the order-by-disorder approach is an intriguing avenue for further study.

The notion that fluctuations in the direction of the spiral order might drive the new physics in MnSi has been suggested by several authors 
\cite{Vojta+01+Schmalian+04,Belitz+06,Tewari+06} working with extensions of Hertz-Millis theory. Indeed, the blue fog phase of Ref. \cite{Tewari+06} is similar in broad 
concept to our picture for the partially ordered phase. The complementary perspective afforded by the quantum order-by-disorder approach --- focussing upon 
particle-hole fluctuations in the presence of spiral order rather than fluctuations of the order parameter --- allows a more tractable 
calculations and leads to some qualitatively different results. We anticipate a similar 
utility in the broader context of phase-formation near to quantum critical points.

\textbf{Acknowledgment:}  The authors benefited from stimulating discussions with 
 D. Belitz, G.~J. Conduit, A.~P. Mackenzie,  A. Rosch, B.~D. Simons and J. Zaanen. We thank C. Pfleiderer for comments on the manuscript 
 and allowing us to use his data. This work was supported by EPSRC under grant code EP/I 004831/1.


\begin{thebibliography}{30}
\expandafter\ifx\csname natexlab\endcsname\relax\def\natexlab#1{#1}\fi
\expandafter\ifx\csname bibnamefont\endcsname\relax
  \def\bibnamefont#1{#1}\fi
\expandafter\ifx\csname bibfnamefont\endcsname\relax
  \def\bibfnamefont#1{#1}\fi
\expandafter\ifx\csname citenamefont\endcsname\relax
  \def\citenamefont#1{#1}\fi
\expandafter\ifx\csname url\endcsname\relax
  \def\url#1{\texttt{#1}}\fi
\expandafter\ifx\csname urlprefix\endcsname\relax\def\urlprefix{URL }\fi
\providecommand{\bibinfo}[2]{#2}
\providecommand{\eprint}[2][]{\url{#2}}

\bibitem[{\citenamefont{Bak and Jensen}(1980)}]{Bak+80}
\bibinfo{author}{\bibfnamefont{P.}~\bibnamefont{Bak}} \bibnamefont{and}
  \bibinfo{author}{\bibfnamefont{H.~H.} \bibnamefont{Jensen}},
  \bibinfo{journal}{J. Phys. C: Solid St. Phys.} \textbf{\bibinfo{volume}{13}},
  \bibinfo{pages}{L881} (\bibinfo{year}{1980}).

\bibitem[{\citenamefont{Ishikawa et~al.}(1976)\citenamefont{Ishikawa, Tajima,
  Bloch, and Roth}}]{Ishikawa+76}
\bibinfo{author}{\bibfnamefont{Y.}~\bibnamefont{Ishikawa}},  \bibnamefont{\emph{et~al.}},
  \bibinfo{journal}{Solid State Commun.} \textbf{\bibinfo{volume}{19}},
  \bibinfo{pages}{525} (\bibinfo{year}{1976}).

\bibitem[{\citenamefont{Pfleiderer et~al.}(2001)\citenamefont{Pfleiderer,
  Julian, and Lonzarich}}]{Pfleiderer+01}
\bibinfo{author}{\bibfnamefont{C.}~\bibnamefont{Pfleiderer}},
  \bibinfo{author}{\bibfnamefont{S.~R.} \bibnamefont{Julian}},
  \bibnamefont{and} \bibinfo{author}{\bibfnamefont{G.~G.}
  \bibnamefont{Lonzarich}}, \bibinfo{journal}{Nature}
  \textbf{\bibinfo{volume}{414}}, \bibinfo{pages}{427} (\bibinfo{year}{2001});
\bibinfo{author}{\bibfnamefont{N.}~\bibnamefont{Doiron-Leyraud}}, \bibnamefont{\emph{et~al.}},
 \bibinfo{journal}{Nature}
  \textbf{\bibinfo{volume}{425}}, \bibinfo{pages}{595} (\bibinfo{year}{2001}).

\bibitem[{\citenamefont{Pfleiderer et~al.}(2004)\citenamefont{Pfleiderer,
  Reznik, Pintschovius, v.~L{\"o}hneysen, Garst, and Rosch}}]{Pfleiderer+04}
\bibinfo{author}{\bibfnamefont{C.}~\bibnamefont{Pfleiderer}}, \bibnamefont{\emph{et~al.}},
  \bibinfo{journal}{Nature} \textbf{\bibinfo{volume}{427}},
  \bibinfo{pages}{227} (\bibinfo{year}{2004}).

\bibitem[{\citenamefont{Abrikosov and Khalatnikov}(1958)}]{Abrikosov+58}
\bibinfo{author}{\bibfnamefont{A.~A.} \bibnamefont{Abrikosov}}
  \bibnamefont{and} \bibinfo{author}{\bibfnamefont{I.~M.}
  \bibnamefont{Khalatnikov}}, \bibinfo{journal}{Soviet Phys. JETP}
  \textbf{\bibinfo{volume}{6}}, \bibinfo{pages}{888} (\bibinfo{year}{1958});
\bibinfo{author}{\bibfnamefont{R.~A.} \bibnamefont{Duine}} \bibnamefont{and}
  \bibinfo{author}{\bibfnamefont{A.~H.} \bibnamefont{MacDonald}},
  \bibinfo{journal}{Phys. Rev. Lett.} \textbf{\bibinfo{volume}{95}},
  \bibinfo{pages}{230403} (\bibinfo{year}{2005}).

\bibitem[{\citenamefont{Conduit and Simons}(2009)}]{Conduit+09b}
\bibinfo{author}{\bibfnamefont{G.~J.} \bibnamefont{Conduit}} \bibnamefont{and}
  \bibinfo{author}{\bibfnamefont{B.~D.} \bibnamefont{Simons}},
  \bibinfo{journal}{Phys. Rev. A} \textbf{\bibinfo{volume}{79}},
  \bibinfo{pages}{053606} (\bibinfo{year}{2009}).

\bibitem[{\citenamefont{Belitz et~al.}(1997)\citenamefont{Belitz, Kirkpatrick,
  and Vojta}}]{Belitz+97}
\bibinfo{author}{\bibfnamefont{D.}~\bibnamefont{Belitz}},
  \bibinfo{author}{\bibfnamefont{T.~R.} \bibnamefont{Kirkpatrick}},
  \bibnamefont{and} \bibinfo{author}{\bibfnamefont{T.}~\bibnamefont{Vojta}},
  \bibinfo{journal}{Phys. Rev. B} \textbf{\bibinfo{volume}{55}},
  \bibinfo{pages}{9452} (\bibinfo{year}{1997});
\bibinfo{author}{\bibfnamefont{J.}~\bibnamefont{Rech}},
  \bibinfo{author}{\bibfnamefont{C.}~\bibnamefont{P\'epin}}, \bibnamefont{and}
  \bibinfo{author}{\bibfnamefont{A.~V.} \bibnamefont{Chubukov}},
  \bibinfo{journal}{Phys. Rev. B} \textbf{\bibinfo{volume}{74}},
  \bibinfo{pages}{195126} (\bibinfo{year}{2006});
\bibinfo{author}{\bibfnamefont{D.~V.} \bibnamefont{Efremov}},
  \bibinfo{author}{\bibfnamefont{J.~J.} \bibnamefont{Betouras}},
  \bibnamefont{and} \bibinfo{author}{\bibfnamefont{A.}~\bibnamefont{Chubukov}},
  \bibinfo{journal}{Phys. Rev. B} \textbf{\bibinfo{volume}{77}},
  \bibinfo{pages}{220401(R)} (\bibinfo{year}{2008});
\bibinfo{author}{\bibfnamefont{D.~L.} \bibnamefont{Maslov}} \bibnamefont{and}
  \bibinfo{author}{\bibfnamefont{A.~V.} \bibnamefont{Chubukov}},
  \bibinfo{journal}{Phys. Rev. B} \textbf{\bibinfo{volume}{79}},
  \bibinfo{pages}{075112} (\bibinfo{year}{2009}).

\bibitem[{\citenamefont{Hertz}(1976)}]{Hertz76}
\bibinfo{author}{\bibfnamefont{J.~A.} \bibnamefont{Hertz}},
  \bibinfo{journal}{Phys. Rev. B} \textbf{\bibinfo{volume}{14}},
  \bibinfo{pages}{1165} (\bibinfo{year}{1976});
\bibinfo{author}{\bibfnamefont{A.~J.} \bibnamefont{Millis}},
  \bibinfo{journal}{Phys. Rev. B} \textbf{\bibinfo{volume}{48}},
  \bibinfo{pages}{7183} (\bibinfo{year}{1993}).

\bibitem[{\citenamefont{Vojta and Sknepnek}(2001)\citenamefont{Schmalian and Turlakov}(2001)}]{Vojta+01+Schmalian+04}
\bibinfo{author}{\bibfnamefont{T.}~\bibnamefont{Vojta}} \bibnamefont{and}
  \bibinfo{author}{\bibfnamefont{R.}~\bibnamefont{Sknepnek}},
  \bibinfo{journal}{Phys. Rev. B} \textbf{\bibinfo{volume}{64}},
  \bibinfo{pages}{052404} (\bibinfo{year}{2001});
\bibinfo{author}{\bibfnamefont{J.}~\bibnamefont{Schmalian}} \bibnamefont{and}
  \bibinfo{author}{\bibfnamefont{M.}~\bibnamefont{Turlakov}},
  \bibinfo{journal}{Phys. Rev. Lett.} \textbf{\bibinfo{volume}{93}},
  \bibinfo{pages}{036405} (\bibinfo{year}{2001}).

\bibitem[{\citenamefont{Pfleiderer et~al.}(1997)\citenamefont{Pfleiderer,
  McMullan, Julian, and Lonzarich}}]{Pfleiderer+97}
\bibinfo{author}{\bibfnamefont{C.}~\bibnamefont{Pfleiderer}}, \bibnamefont{\emph{et~al.}},
 \bibinfo{journal}{Phys. Rev. B}
  \textbf{\bibinfo{volume}{55}}, \bibinfo{pages}{8330} (\bibinfo{year}{1997}).

\bibitem[{\citenamefont{Uhlarz et~al.}(2004)\citenamefont{Uhlarz, Pfleiderer,
  and Hayden}}]{Uhlarz+04}
\bibinfo{author}{\bibfnamefont{M.}~\bibnamefont{Uhlarz}},
  \bibinfo{author}{\bibfnamefont{C.}~\bibnamefont{Pfleiderer}},
  \bibnamefont{and} \bibinfo{author}{\bibfnamefont{S.~M.}
  \bibnamefont{Hayden}}, \bibinfo{journal}{Phys. Rev. Lett.}
  \textbf{\bibinfo{volume}{93}}, \bibinfo{pages}{256404}
  (\bibinfo{year}{2004});
\bibinfo{author}{\bibfnamefont{A.}~\bibnamefont{Huxley}}, \bibnamefont{\emph{et~al.}},
  \bibinfo{journal}{Phys. Rev. B} \textbf{\bibinfo{volume}{63}},
  \bibinfo{pages}{144519} (\bibinfo{year}{2002}).

\bibitem[{\citenamefont{Borzi et~al.}(2007)\citenamefont{Borzi, Grigera,
  Farrell, Perry, Lister, Lee, Tennant, Maeno, and Mackenzie}}]{Borzi+07}
\bibinfo{author}{\bibfnamefont{R.~A.} \bibnamefont{Borzi}}, \bibnamefont{\emph{et~al.}},
  \bibinfo{journal}{Science} \textbf{\bibinfo{volume}{315}},
  \bibinfo{pages}{214} (\bibinfo{year}{2007}).

\bibitem[{\citenamefont{Laughlin et~al.}(2001)\citenamefont{Laughlin,
  Lonzarich, Monthoux, and Pines}}]{Laughlin+01}
\bibinfo{author}{\bibfnamefont{R.~B.} \bibnamefont{Laughlin}}, \bibnamefont{\emph{et~al.}},
  \bibinfo{journal}{Adv. Phys.} \textbf{\bibinfo{volume}{50}},
  \bibinfo{pages}{361} (\bibinfo{year}{2001}).

\bibitem[{\citenamefont{Conduit et~al.}(2009)\citenamefont{Conduit, Green, and
  Simons}}]{Conduit+09}
\bibinfo{author}{\bibfnamefont{G.~J.} \bibnamefont{Conduit}},
  \bibinfo{author}{\bibfnamefont{A.~G.} \bibnamefont{Green}}, \bibnamefont{and}
  \bibinfo{author}{\bibfnamefont{B.~D.} \bibnamefont{Simons}},
  \bibinfo{journal}{Phys. Rev. Lett.} \textbf{\bibinfo{volume}{103}},
  \bibinfo{pages}{207201} (\bibinfo{year}{2009}).

\bibitem[{\citenamefont{Mila et~al.}(1991)\citenamefont{Mila, Poilblanc, and
  Bruder}}]{Mila+91}
\bibinfo{author}{\bibfnamefont{F.}~\bibnamefont{Mila}},
  \bibinfo{author}{\bibfnamefont{D.}~\bibnamefont{Poilblanc}},
  \bibnamefont{and} \bibinfo{author}{\bibfnamefont{C.}~\bibnamefont{Bruder}},
  \bibinfo{journal}{Phys. Rev. B} \textbf{\bibinfo{volume}{43}},
  \bibinfo{pages}{789} (\bibinfo{year}{1991});
\bibinfo{author}{\bibfnamefont{J.}~\bibnamefont{Zaanen}},
  \bibinfo{journal}{Phys. Rev. Lett.} \textbf{\bibinfo{volume}{84}},
  \bibinfo{pages}{753} (\bibinfo{year}{2000});
\bibinfo{author}{\bibfnamefont{J.}~\bibnamefont{Zaanen}}, \bibnamefont{\emph{et~al.}},
  \bibinfo{journal}{Philos. Mag. B} \textbf{\bibinfo{volume}{81}},
  \bibinfo{pages}{1485} (\bibinfo{year}{2001});
\bibinfo{author}{\bibfnamefont{F.}~\bibnamefont{Kr\"uger}} \bibnamefont{and}
  \bibinfo{author}{\bibfnamefont{S.}~\bibnamefont{Scheidl}},
  \bibinfo{journal}{Europhys. Lett.} \textbf{\bibinfo{volume}{74}},
  \bibinfo{pages}{896} (\bibinfo{year}{2006}).


\bibitem[{\citenamefont{Frigeri et~al.}(2004)\citenamefont{Frigeri, Agterberg,
  Koga, and Sigrist}}]{Frigeri+04}
\bibinfo{author}{\bibfnamefont{P.~A.} \bibnamefont{Frigeri}}, \bibnamefont{\emph{et~al.}},
  \bibinfo{journal}{Phys. Rev. Lett.} \textbf{\bibinfo{volume}{92}},
  \bibinfo{pages}{097001} (\bibinfo{year}{2004}).

\bibitem[{\citenamefont{Belitz et~al.}(2006)\citenamefont{Belitz, Kirkpatrick,
  and Rosch}}]{Belitz+06}
\bibinfo{author}{\bibfnamefont{D.}~\bibnamefont{Belitz}},
  \bibinfo{author}{\bibfnamefont{T.~R.} \bibnamefont{Kirkpatrick}},
  \bibnamefont{and} \bibinfo{author}{\bibfnamefont{A.}~\bibnamefont{Rosch}},
  \bibinfo{journal}{Phys. Rev. B} \textbf{\bibinfo{volume}{73}},
  \bibinfo{pages}{054431} (\bibinfo{year}{2006}).

\bibitem[{\citenamefont{Karahasanovic et~al.}()\citenamefont{Karahasanovic,
  Kr\"uger, and Green}}]{Una_unpublished}
\bibinfo{author}{\bibfnamefont{U.}~\bibnamefont{Karahasanovic}},
  \bibinfo{author}{\bibfnamefont{F.}~\bibnamefont{Kr\"uger}}, \bibnamefont{and}
  \bibinfo{author}{\bibfnamefont{A.~G.} \bibnamefont{Green}},
  \emph{\bibinfo{title}{In preparation}}.
  
\bibitem[{not({\natexlab{b}})}]{note}
\bibinfo{note}{The $\ln T$-dependence of $b_0$ reflects the $M^4 \ln [M^2+T^2]$ in extensions to Hertz-Millis theory.}

  
\bibitem[{\citenamefont{Binz et~al.}(2006)\citenamefont{Binz, Vishwanath, and
  Aji}}]{Binz+06}
\bibinfo{author}{\bibfnamefont{B.}~\bibnamefont{Binz}},
  \bibinfo{author}{\bibfnamefont{A.}~\bibnamefont{Vishwanath}},
  \bibnamefont{and} \bibinfo{author}{\bibfnamefont{V.}~\bibnamefont{Aji}},
  \bibinfo{journal}{Phys. Rev. Lett.} \textbf{\bibinfo{volume}{96}},
  \bibinfo{pages}{207202} (\bibinfo{year}{2006});
\bibinfo{author}{\bibfnamefont{U.~K.} \bibnamefont{Roszler}},
  \bibinfo{author}{\bibfnamefont{A.~N.} \bibnamefont{Bogdanov}},
  \bibnamefont{and}
  \bibinfo{author}{\bibfnamefont{C.}~\bibnamefont{Pfleiderer}},
  \bibinfo{journal}{Nature} \textbf{\bibinfo{volume}{442}},
  \bibinfo{pages}{797} (\bibinfo{year}{2006});
\bibinfo{author}{\bibfnamefont{I.}~\bibnamefont{Fischer}},
  \bibinfo{author}{\bibfnamefont{N.}~\bibnamefont{Shah}}, \bibnamefont{and}
  \bibinfo{author}{\bibfnamefont{A.}~\bibnamefont{Rosch}},
  \bibinfo{journal}{Physical Review B} \textbf{\bibinfo{volume}{77}},
  \bibinfo{pages}{024415} (\bibinfo{year}{2008}).

\bibitem[{\citenamefont{Neubauer et~al.}(2009)\citenamefont{Neubauer,
  Pfleiderer, Binz, Rosch, Ritz, Niklowitz, and B\"oni}}]{Neubauer+98}
\bibinfo{author}{\bibfnamefont{A.}~\bibnamefont{Neubauer}}, \bibnamefont{\emph{et~al.}},
  \bibinfo{journal}{Phys. Rev. Lett.} \textbf{\bibinfo{volume}{102}},
  \bibinfo{pages}{186602} (\bibinfo{year}{2009}).

\bibitem[{\citenamefont{Tewari et~al.}(2006)\citenamefont{Tewari, Belitz, and
  Kirkpatrick}}]{Tewari+06}
\bibinfo{author}{\bibfnamefont{S.}~\bibnamefont{Tewari}},
  \bibinfo{author}{\bibfnamefont{D.}~\bibnamefont{Belitz}}, \bibnamefont{and}
  \bibinfo{author}{\bibfnamefont{T.~R.} \bibnamefont{Kirkpatrick}},
  \bibinfo{journal}{Phys. Rev. Lett.} \textbf{\bibinfo{volume}{96}},
  \bibinfo{pages}{047207} (\bibinfo{year}{2006}).

\bibitem[{\citenamefont{Kirkpatrick and Belitz}(2010)}]{Kirkpatrick+10}
\bibinfo{author}{\bibfnamefont{T.~R.} \bibnamefont{Kirkpatrick}}
  \bibnamefont{and} \bibinfo{author}{\bibfnamefont{D.}~\bibnamefont{Belitz}},
  \bibinfo{journal}{Phys. Rev. Lett.} \textbf{\bibinfo{volume}{104}},
  \bibinfo{pages}{256404} (\bibinfo{year}{2010}).

\end{thebibliography}
\end{document}